\documentclass[
    reprint,
    preprintnumbers,
    superscriptaddress,
    nofootinbib,
     amsmath,amssymb,
     aps,
     prd,
    floatfix,
    ]{revtex4-2}

    \usepackage{graphicx}
    \usepackage[caption=false]{subfig}
    \usepackage[usenames,dvipsnames]{xcolor} % colour
    \usepackage[utf8]{inputenc}
    \usepackage{color}
    \usepackage{hyperref}
    \usepackage[normalem]{ulem} 
    \usepackage{physics}

    \usepackage{enumitem}
    \usepackage{comment}
    \usepackage{bm}
\hypersetup{
  colorlinks   = true, %Colours links instead of ugly boxes
  urlcolor     = blue, %Colour for external hyperlinks
  linkcolor    = blue, %Colour of internal links
  citecolor   = red %Colour of citations
}
    \allowdisplaybreaks[1]
    \graphicspath{{figures/}}

\usepackage{float}

\newcommand{\mpl}{M_\mathrm{Pl}}
\newcommand{\reh}{_\mathrm{reh}}
\newcommand{\rhoc}{\rho_\mathrm{c}}
\newcommand{\rhocdec}{\rho^\mathrm{c}_\mathrm{dec}}
\newcommand{\rhocosc}{\rho^\mathrm{c}_\mathrm{osc}}

\newcommand{\adec}{a_\mathrm{dec}}
\newcommand{\deltaini}{\delta_\mathrm{reh}}

\newcommand{\rhorehav}{\bar{\rho}_\mathrm{reh}}

\newcommand{\deltac}{\delta_\mathrm{c}}

\newcommand{\ee}[1]{{\times 10^{#1}}}

\newcommand{\oxford}{Astrophysics, University of Oxford, DWB, Keble Road, Oxford OX1 3RH, UK}
\newcommand{\qmul}{Geometry, Analysis and Gravitation, School of Mathematical Sciences, Queen Mary University of London,
Mile End Road, London E1 4NS, UK}
\newcommand{\damtp}{DAMTP, Centre for Mathematical Sciences, Wilberforce Road, Cambridge, CB3 0WA, UK}

\begin{document}

\title{Oscillon formation during inflationary preheating with general relativity}

\author{Josu C. Aurrekoetxea}
\email{josu.aurrekoetxea@physics.ox.ac.uk}
\affiliation{\oxford}
\author{Katy Clough}
\email{k.clough@qmul.ac.uk}
\affiliation{\qmul}
\author{Francesco Muia}
\email{fm538@cam.ac.uk}
\affiliation{\damtp}

\begin{abstract}
We study the non-perturbative evolution of inflationary fluctuations during preheating using fully non-linear general-relativistic field-theory simulations. We choose a single-field inflationary model that is consistent with observational constraints and start the simulations at the end of inflation with fluctuations both in the field and its conjugate momentum. Gravity enhances the growth of density perturbations, which then collapse and virialize, forming long-lived stable oscillon-like stars that reach compactnesses $\mathcal{C}\equiv GM/R \sim 10^{-3}-10^{-2}$. We find that $\mathcal{C}$ increases for larger field models, until it peaks due to the interplay between the overdensity growth and Hubble expansion rates. Whilst gravitational effects can play an important role in the formation of compact oscillons during preheating, the objects are unlikely to collapse into primordial black holes without an additional enhancement of the initial inflationary fluctuations.
\end{abstract}
\keywords{Inflation, Preheating, Oscillons, General Relativity, Numerical Relativity, Simulations}

\maketitle

%%%%%%%%%%%%%%%%%%%%%%%%%%%%%%%%%%%%%%%%%%%%%%%%%%%%%%%%%%%%%%

\section{Introduction}

Cosmic inflation \cite{Guth:1980zm,Starobinsky:1980te,Linde:1981mu,Albrecht:1982wi} is a period of accelerated expansion of the very early Universe that solves several puzzles in the standard hot Big Bang theory and provides an elegant mechanism for the production of the anisotropies observed in the Cosmic Microwave Background (CMB)\cite{Planck:2018vyg}. In the simplest models, the accelerated expansion is driven by a single slowly-rolling scalar field $\phi$, called the \textit{inflaton}. Current constraints \cite{Planck:2018jri} favour plateau-like potentials that open-up away from the minimum where inflation ends, such as the so-called $\alpha$-attractor models \cite{Kallosh:2013hoa,Kallosh:2013yoa}
\begin{equation}\label{eq:potential}
V(\phi) = \frac{m^2 \mu^2}{2} \left(1 - e^{\phi/\mu} \right)^2 ,
\end{equation}
where $\mu$ can vary over a wide range of scales and parameterises whether the potential is small field ($\mu \ll \mpl $) or large field ($\mu \sim \mpl $).
One poorly understood aspect of the early Universe is the period that connects inflation to Big Bang Nucleosynthesis (BBN). In particular, the details of \textit{reheating} \cite{Kofman:1994rk, Kofman:1997yn,Albrecht:1982mp}, the mechanism by which the energy density stored in the inflaton is transferred into the Standard Model (SM) sector, are unclear. The simplest channel is through the perturbative decay of the inflaton $\phi$ to SM particle(s) $\psi$, e.g. mediated by $\propto \phi \bar\psi \psi$ couplings in the Lagrangian.

\begin{figure}[t!]
    \centering
    \includegraphics[width=\linewidth]{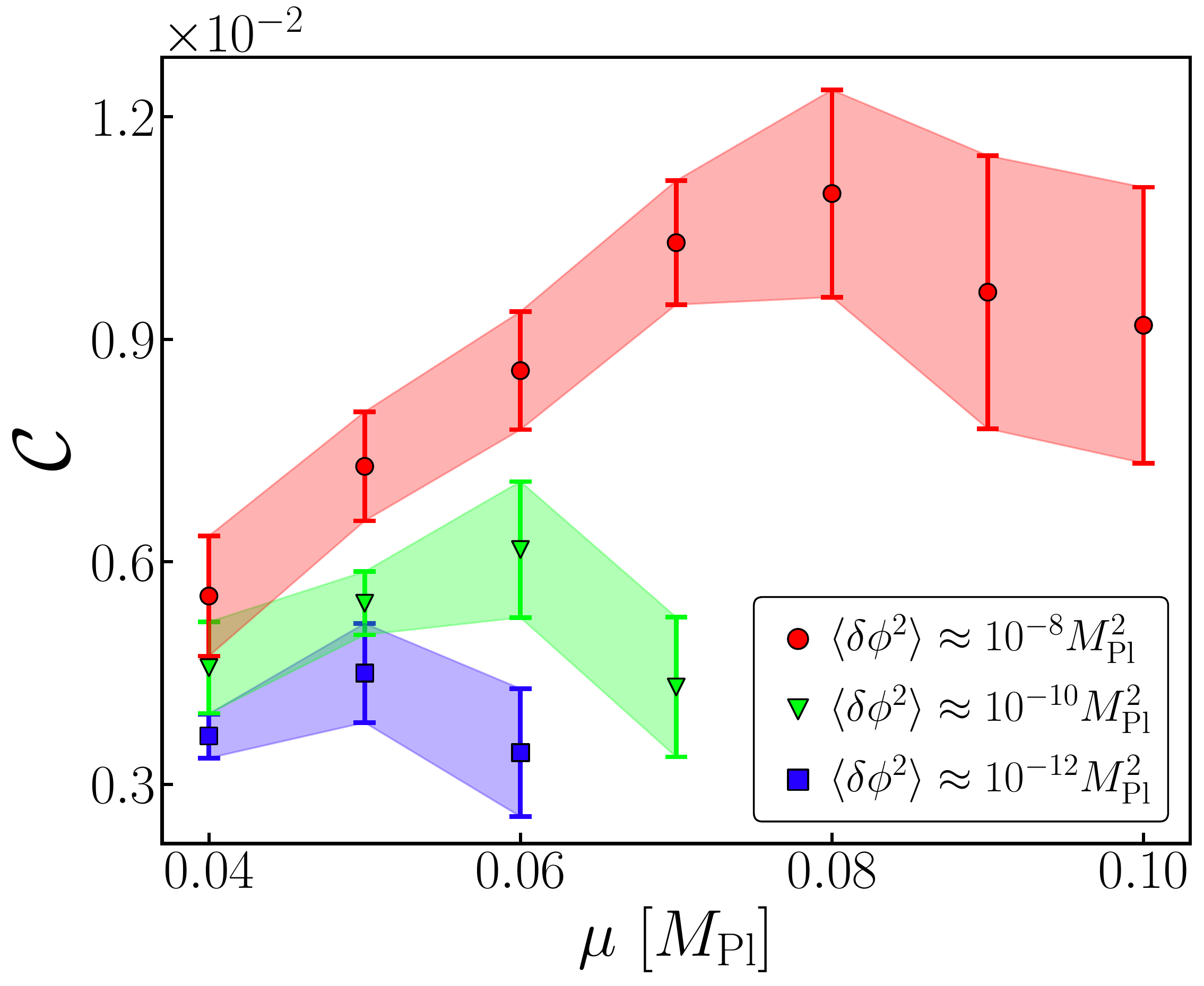}
\caption{Compactness $\mathcal{C}=G M/R$ of oscillons formed during inflationary preheating for different scales of the potential $\mu$ and initial amplitude fluctuations $\langle\delta\phi^2\rangle$.
The interplay between the overdensity growth and Hubble rate results in a maximum compactness for each combination $(\mu,\langle \delta\phi^2 \rangle)$.}
\label{fig:oscillon}
\end{figure}
\begin{figure*}[t!]
    \centering
    \href{https://youtu.be/vTl9agMfPB0}{
    \includegraphics[width=\linewidth]{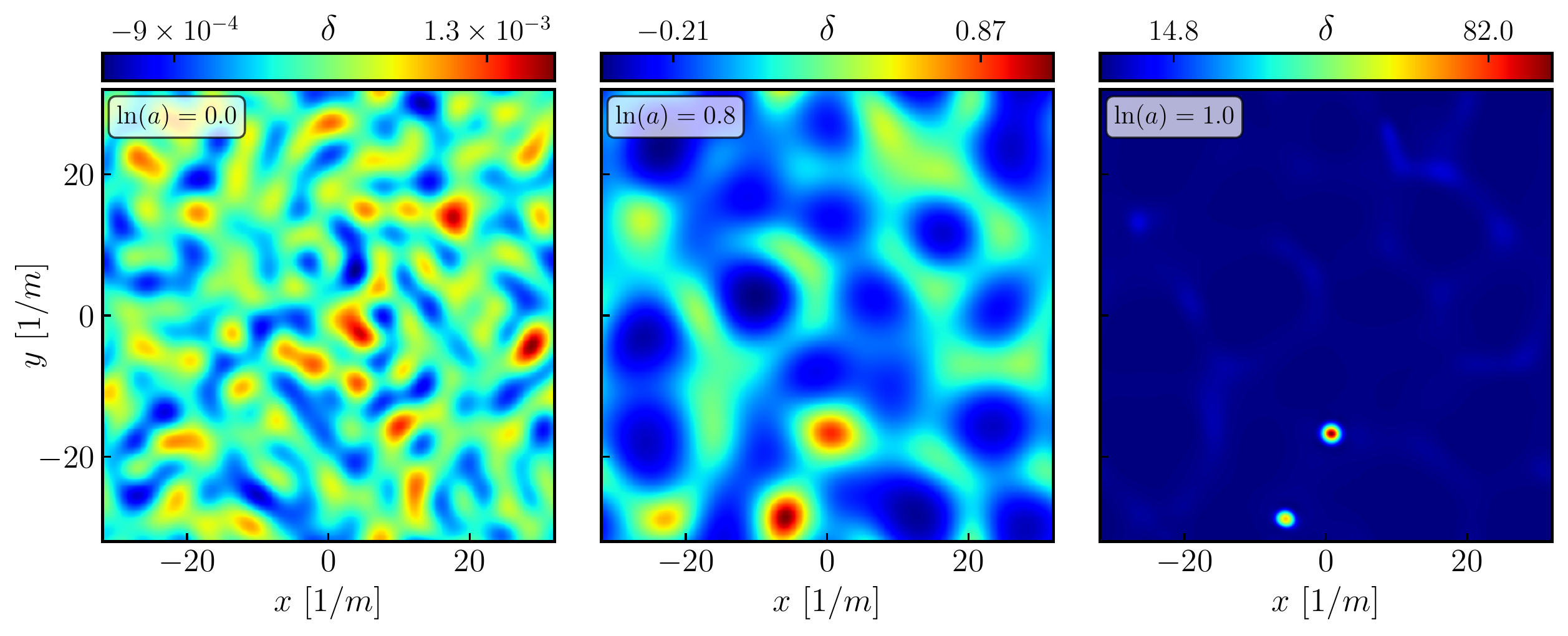}}
\caption{Non-perturbative evolution of inflationary fluctuations during preheating. We plot the spatial slices of the density contrast $\delta\equiv \rho/\bar{\rho} - 1$ in our simulation coordinates at three e-foldings $\ln(a) = \lbrace 0, 0.8, 1.0\rbrace$, where $a$ is the spatially-averaged scale factor. Overdensities grow rapidly and collapse into stable oscillon-like objects. Movie: \url{https://youtu.be/vTl9agMfPB0}.} 
\label{fig:panel}
\end{figure*}

It is possible that some additional non-perturbative dynamics take place between the end of inflation and reheating, featuring resonances that lead to exponentially growing solutions for $\phi$. This process, known as \textit{preheating} \cite{Amin:2014eta}, has been extensively studied using both analytical and numerical techniques \cite{Amin:2010jq, Amin:2010dc, Amin:2011hj, Amin:2013ika, Felder:2006cc, Lozanov:2014zfa, Antusch:2015ziz, Antusch:2015vna, DeCross:2015uza,DeCross:2016fdz,DeCross:2016cbs,Kim:2017duj,Lozanov:2017hjm, Musoke:2019ima, Niemeyer:2019gab,Nguyen:2019kbm,Martin:2019nuw,Eggemeier:2020zeg,Iarygina:2020dwe,Sang:2020kpd,Martin:2020fgl,vandeVis:2020qcp,Kost:2021rbi,Garcia:2021iag,Eggemeier:2021smj,Kim:2021ipz,Figueroa:2022iho,Mahbub:2023faw}. These large fluctuations in the field can collapse into stable scalar field configurations known as \textit{oscillons}\footnote{We will collectively use the term \textit{oscillon} to denote all real scalar field pseudo-stable compact objects.} \cite{Bogolyubsky:1976nx,Bogolyubsky:1976sc,Gleiser:1993pt,Copeland:1995fq,Kasuya:2002zs,Saffin:2006yk,Hertzberg:2010yz,Salmi:2012ta,Gleiser:2019rvw,Antusch:2019qrr,Ibe:2019vyo,Zhang:2020bec,vanDissel:2023zva} and source a stochastic background of gravitational waves (GWs) \cite{Caprini:2018mtu,Dufaux:2007pt,Zhou:2013tsa,Antusch:2016con,Figueroa:2016ojl,Antusch:2017vga,Figueroa:2017vfa,Amin:2018xfe,Armendariz-Picon:2019csc,Armendariz-Picon:2020tkc,Hiramatsu:2020obh,Bhoonah:2020oov,Cai:2021gju,Adshead:2019lbr,Sang:2019ndv,Cui:2021are,Cosme:2022htl,Eggemeier:2022gyo,Krajewski:2022ezo,Lozanov:2022yoy}, which could be probed by future detectors targeting the $\mathrm{MHz-GHz}$ frequency band~\cite{Aggarwal:2020olq}. Given that the Universe is opaque to light before the release of the CMB photons, GWs might be the only way to directly probe such early epochs.

Our goal in this paper is to study to what extent gravity plays a role in the formation of (compact) oscillons during preheating, and to quantify the maximum compactness that can be achieved. Our results are summarised in Fig. \ref{fig:oscillon}, where we see that the interplay between the overdensity growth and Hubble rate results in a maximum compactness for each combination $(\mu,\langle \delta\phi^2 \rangle)$. We focus on what is (arguably) the minimal case -- when the only field involved is the inflaton -- and investigate scales of the potential $\mu$, with self-consistent fluctuations $\delta\phi(t,\mathbf{x})$ in both the field and momentum that recover values of the spectral index and tensor-to-scalar ratio that are allowed by current Planck bounds \cite{Planck:2018jri}\footnote{Preheating could also involve an extra field that is not responsible for inflation, e.g. a string modulus \cite{Antusch:2017vga}, in which case parameters are not dictated by the observational constraints on inflation.}.

Preheating has been extensively studied using lattice field theory simulations \cite{Felder:2000hq,Frolov:2008hy,Figueroa:2020rrl, Figueroa:2021yhd} that evolve the scalar field equations on a homogeneous Friedmann-Lema{\^ i}tre-Robertson-Walker (FLRW) background, and neglect the backreaction of inhomogeneities on the local spacetime metric. These effects become important when overdensities grow large enough for gravity to be of the same order as self interactions in the field. As a first approximation, this can be accounted for by evolving a Newtonian potential with a Poisson equation sourced by the energy density~\cite{Lozanov:2019ylm, Amin:2019ums}.
Recently, a few works \cite{Kou:2019bbc, Giblin:2019nuv, Kou:2021bij} included the effects of the gravitational backreaction at the fully non-linear level of general relativity showing interesting phenomenological effects, that deviated from the evolution in a spatially constant expanding metric background.

In this paper we use numerical relativity to study a simple inflationary model with consistent fluctuations in both the scalar field and conjugate momentum at the end of inflation. In particular, we aim to quantify the compactnesses of the oscillons that are formed for different scales $\mu$ of the potential. For larger $\mu$, gravity enhances the growth of small overdensities during inflationary preheating, which then decouple from the Hubble flow, collapse and virialize, forming stable long-lived oscillon-like stars that reach compactnesses $\mathcal{C}\equiv GM/R \sim 10^{-3}-10^{-2}$.  We find that $\mathcal{C}$ increases with the potential scale $\mu$, until it reaches a maximum value due to a balance between the overdensity growth rate and dilution due to the universe's expansion rate. This maximum compactness depends on the initial amplitude of the fluctuations and for $\langle\delta\phi^2\rangle \approx 10^{-8}\mpl^2$ is $\mathrm{max}(\mathcal{C})\approx 10^{-2}$. These compactnesses result in a significant backreaction on the local spacetime metric that may have implications for the amplitude of the gravitational-wave background but they are insufficient to collapse the objects into primordial black holes.\\

The paper is structured as follows: in Sec. \ref{sec:methods} we describe the key points of the methods used to set up and simulate the post-inflationary spacetime and track the evolution of perturbations. In Sec. \ref{sec:results-growth} we characterise the non-linear growth of the overdensities for different initial amplitudes $\langle \delta\phi^2 \rangle$ and potential scales $\mu$. In Sec. \ref{sec:results-endpoint} we describe their endpoint as stable gravitationally bound objects and comment on their masses, radii and compactnesses. We conclude and suggest future directions in Sec. \ref{sec:conclusion}. Additional details of the numerics and testing are included in the Appendices \ref{app:numerics}, \ref{app:FLRW} and \ref{app:testing}.

\newpage

\section{Simulating inflationary preheating}
\label{sec:methods}

The simplest single-field inflationary mechanism can be generated via the action \begin{equation}
    S = \int \dd^4 x \sqrt{-g}\left(\frac{\mpl^2}{2}R - \frac{1}{2}\nabla_{\mu}\phi\nabla^{\mu}\phi - V(\phi)\right)\,,
\end{equation}
where $\mpl = \sqrt{1/8\pi G}$ is the reduced Planck mass. The evolution of the scalar field is given by the Klein-Gordon equation
\begin{equation}\label{eq:EKG_cov}
    \nabla^\mu \nabla_\mu \phi - V'(\phi) = 0~,
\end{equation}
where $V(\phi)$ is the inflationary potential. The preheating mechanism can be understood by splitting the inflaton field in a homogeneous component and a perturbation, $\phi(t, \mathbf{x}) = \bar{\phi}(t) + \delta\phi(t, \mathbf{x})$. While the field is oscillating around the minimum of its potential, %after the end of inflation, 
the equation of motion for the perturbation mode $\delta\phi_k(t)$ (in Fourier space) becomes
\begin{equation}
    \delta\ddot\phi_k + \left(k^2 + V''\left(\phi(t)\right)\right) \delta\phi_k(t) = 0  \,,
    \label{eq:PerturbationsEOM}
\end{equation}
where the prime and dot denote differentiation with respect to the field $\phi$ and time, respectively. Here $\mathbf{k}$ is the wavemode considered and $k = |\mathbf{k}|$. Depending on the wavemode and details of $V(\phi)$, the equations of motion can feature resonances that lead to exponentially growing solutions for $\delta\phi$:
\begin{itemize}[leftmargin=*]
\item If $\bar{\phi}(t)$ is a periodic function of time, $V(\phi)$ can act as driving force leading to a phenomenon called \textit{parametric resonance}. 
\item If $\bar{\phi}(t)$ repeatedly probes regions of the scalar potential where $V'' < 0$, perturbations can grow exponentially via \textit{tachyonic resonance}. 
\end{itemize}

In this work we will study these resonances within the inflaton field for the $\alpha$-attractor model in Eqn. \eqref{eq:potential}, where $\mu$ can vary over a wide range of scales\footnote{The Starobinsky model is a particular example where the scale $\mu$ and the Planck scale share a common origin $\mu = \sqrt{3/2}\mpl$.} and parameterises whether the potential is small or large field, that is, how far in field space is the inflationary region from the reheating minimum, compared to $\mpl$. Near the reheating minimum
\begin{equation}\label{eq:pot_expansion}
    V(\phi)\approx \frac{m^2\phi^2}{2}\left(1 + \frac{\phi}{\mu} + \frac{7}{12}\frac{\phi^2}{\mu^2}\right) + \mathcal{O}(\phi^5) \,.
\end{equation}
The average field value $\bar\phi(t)$ oscillates around the minimum at $\phi = 0$, giving rise to a matter dominated era. The odd powers of $\phi$ in the expansion in Eqn.~\eqref{eq:pot_expansion} ensure that the potential is shallower than quadratic for $\phi < 0$, so that $\phi$ particles feel an attractive force when the average field value is probing that region of the potential. Such an attractive force can stabilise overdensities in the field against dispersion, allowing the existence of oscillon-like solutions.\\

To describe the metric sector, we decompose the four-dimensional line element into the 3+1D ADM form
\begin{equation}
    ds^2 = -\alpha^2dt^2 + \gamma_{ij}(dx^i + \beta^i dt)(dx^j + \beta^j dt),
\end{equation}
where $\alpha$ and $\beta^i$ are the lapse and shift gauge functions, and $\gamma_{ij}$ is the three-dimensional spatial metric. We also evolve the extrinsic curvature tensor $K_{ij}=\partial_{t} \gamma_{ij} + 2 D_{(i} \beta_{j)}$, which can be decomposed into a trace $K=\gamma^{ij}K_{ij}$  and a traceless part $A_{ij}$. In numerical relativity simulations it is common to decompose the spatial metric $\gamma_{ij}=\chi^{-1}\tilde{\gamma}_{ij}$ with a conformally related metric $\tilde{\gamma}_{ij}$ that has unit determinant $\tilde{\gamma} = 1$, whilst the traceless part of the extrinsic curvature is similarly rescaled as $\tilde{A}_{ij} = \chi A_{ij}$.  In the FLRW limit, the conformal factor is $\chi^{-1}=a(t)^2$ and the expansion $K$ is related to the Hubble parameter as $K=-3H$. In addition, the traceless part of the extrinsic curvature tensor contains details about the energy density in gravitational waves\footnote{This is equivalent to the energy density of the Isaacson energy momentum tensor $t_{00} = \langle\dot{h}_{\mu\nu} \dot{h}^{\mu\nu}\rangle /32\pi G$ when the perturbations are small and one can average over one period in spacetime \cite{Isaacson:1968hbi}.} as $\rho_\mathrm{GW}\propto \tilde{A}_{ij}\tilde{A}^{ij}$.

\subsection{Initial conditions}

We study different $\alpha$-attractor models parameterised by the scale $\mu$. We set the homogeneous value of the scalar field prior to inflation $\phi_\mathrm{inf}=\phi(t_\mathrm{inf})$ by requiring inflation to last $\mathcal{N}=\ln(a) \approx 50$ e-folds. The mass $m$ is fixed such that fluctuations are consistent with the scalar power spectrum observed in the CMB \cite{Planck:2018jri}
\begin{equation}
    \Delta_{\mathcal R}^2 = \frac{H_\mathrm{inf}^2}{8\pi^2 \mpl^2 \epsilon(\phi_\mathrm{inf})} \approx 2\times 10^{-9} \,,
\end{equation}
where $\mathcal{R}$ is the curvature perturbation.
We solve the ODE for the homogeneous equations of motion until the end of inflation (which corresponds to the beginning of reheating $t_\mathrm{reh}$), see Appendix \ref{app:numerics} for more details. We identify $\phi\reh = \phi(t\reh)$ and $\dot{\phi}\reh= \dot{\phi}(t\reh)$, which we use as the background on which to construct the inhomogeneous initial conditions for our scalar field and momentum
\begin{equation}
    \phi(\mathbf{x}) = \phi\reh + \delta\phi(\mathbf{x}) \qquad \dot{\phi}(\mathbf{x}) = \dot{\phi}\reh + \delta\dot{\phi}(\mathbf{x}) ~.
\end{equation}
Assuming a random Gaussian field, the spectrum of sub-horizon scalar perturbations at the end of inflation is determined by quantum vacuum fluctuations, i.e.
\begin{align}
\label{eq:PertSpectrum_main}
\mathcal{P}(k) &= \frac{\lambda}{2 a_\mathrm{reh}^2 \omega_k^2} \,, \\
\langle \delta\phi_\mathbf{k} \delta\phi_{\mathbf{k}'} \rangle &= (2\pi)^3 \mathcal{P}(k) \delta(\mathbf{k}-\mathbf{k}') \,,
\end{align}
where $\omega_k = \sqrt{k^2 + a_\mathrm{reh}^2 V''(\phi\reh)}$, $\lambda = 1$ and we choose the initial scale factor $a_\mathrm{reh}=a(t_\mathrm{reh})=1$. The spectrum in Eq.~\eqref{eq:PertSpectrum_main} determines the variance of the field perturbations through the relation
\begin{equation}
\langle \delta\phi^2 \rangle = \int d\log k \, \frac{k^3}{2 \pi^2} \mathcal{P}(k) \,,
\end{equation}
which will be one of the quantities varied in our simulation by considering the three cases $\lambda \in \{1, 10^2, 10^4\}$. Note that we are considering modes that are sub-horizon at the end of inflation and never exited the horizon during the inflationary stage.
These modes correspond to physical scales that are not constrained by CMB observations, but one may expect they have standard quantum vacuum fluctuation amplitudes. See Ref. \cite{Figueroa:2020rrl} and Appendix \ref{app:numerics} for more details about the lattice implementation.\\

The fluctuations source both the energy and momentum density components of the stress-energy tensor measured by the normal observers
\begin{align}
    \rho &= \frac{1}{2}(\partial_i \phi)^2 + \frac{1}{2}\dot{\phi}^2 + V(\phi)~, \\
    S_i &= -\dot{\phi} \partial_i \phi~.
\end{align}
We solve both the Hamiltonian and momentum constraints to construct valid initial data for the scalar and gravitational sectors. We formulate the coupled system of non-linear elliptic equations using the CTTK method \cite{Aurrekoetxea:2022mpw}, which assumes an initially conformally flat metric $\tilde{\gamma}_{ij} = \delta_{ij}$ and chooses an initial conformal (or equivalently, scale) factor. In this case we have chosen $\chi$ = 1 initially (equivalent to choosing $a(t\reh)=1$ in the homogeneous case). The method then solves the constraints as an algebraic equation\footnote{Note that in the absence of inhomogeneities, this Hamiltonian constraint in the CTTK method reduces to the usual form of the Friedmann constraint $H^2 = \rho/3\mpl^2$, since $K=-3H$.} and a Poisson-like equation that in the case in which initially $\chi=1$, reduce to
\begin{align}
    K^2 = \frac{3\rho}{\mpl^2} + \frac{3}{2}A_{ij}A^{ij} \,,\\
    \delta^{jk}\partial_k A_{ij} = \frac{2}{3} \partial_i K + \frac{S_i}{\mpl^2}~.
\end{align}
In some previous works $\dot{\phi}$ is set to zero in order to simplify the solution of the constraints. We find that including the homogeneous component of the scalar field momentum $\dot{\phi}\reh$ is crucial, because it allows the field to explore the tachyonic part of the potential during the first few oscillations, which significantly enhances the growth rate of overdensities. However, including fluctuations in the conjugate momenta of the field perturbations -- whilst technically correct -- does not appear to significantly impact the results.

\subsection{Evolution}

We evolve the BSSN formulation of the Einstein equations of general relativity \cite{Nakamura:1987zz,Shibata:1995we,Baumgarte:1998te} using the publicly available numerical relativity (NR) code \textsc{grchombo} \cite{Andrade:2021rbd, Radia:2021smk,Clough:2015sqa}, together with a modified version of the integrated moving puncture gauge \cite{Campanelli:2005dd,Baker:2005vv,Kou:2019bbc,Giblin:2019nuv}
\begin{equation}
    \partial_t \alpha = \frac{\alpha}{2}\left(K-\overline{K}\right) + \beta^i\partial_i \alpha~,
\end{equation}
where $\overline{K}$ is the proper-volume-averaged trace of the extrinsic curvature tensor. This gauge choice approximately ensures that simulation time is identified with the cosmic time coordinate rather than conformal time. For the latter choice (which has been used in previous works on early universe phenomena e.g. \cite{Aurrekoetxea:2019fhr,deJong:2021bbo,Joana:2020rxm,Joana:2022uwc}), the lapse function grows with the expansion, and as a result the timescale of the oscillation of the field around the reheating minimum becomes under resolved as $d\tau = \alpha dt$. Note that in simulations of strongly inhomogeneous spacetimes, gauge dependence of the physical quantities measured is practically unavoidable, so we need to take care in interpreting our results.

We consider the growth of density perturbations for a range of potential scales $\mu$ summarised in table \ref{fig:table}. We choose the size of our numerical domain to be approximately the Hubble length $L\approx H(t_\mathrm{reh})^{-1}$, which is mostly determined by the homogeneous components of the initial conditions. This choice introduces an infrared cutoff scale $k_\mathrm{IR}=2\pi/L$ of the modes that can be studied. Likewise, we enforce an ultraviolet cutoff $k_\mathrm{UV}=4k_\mathrm{IR}$ to ensure that modes in the lattice have good spatial resolution. We fix $L=64 m^{-1}$ for all simulations\footnote{This value corresponds to the largest Hubble length of the studied models (see table \ref{fig:table}), and is thus the most conservative choice to avoid artificial boundary effects (that is, imposing unwanted periodicity on scales that are in causal contact.}, which includes a range of modes $k/m \subset \{0.1,\, 0.4\}$ that experience parametric resonance. We evolve the perturbations until we see that either they collapse into compact objects and stabilise, or the growth saturates without forming compact structures.

\subsection{Diagnostics}

We focus on diagnostics that give us information about the local behaviour of the perturbations, such as the local density contrast 
\begin{equation}
\deltac \equiv \frac{\rhoc}{\bar{\rho}} - 1~,
\end{equation}
where $\rhoc$ is the central value of the overdensities and $\bar{\rho}$ is the mean energy density across the box, which approximately tracks the evolution of a matter-dominated spacetime $\bar{\rho}\sim a^{-3}$ due to the harmonic oscillations in the averaged field. Note that whilst $\deltac$ can be calculated locally it does have a dependence on the global and local slicing (gauge choice) of the spacetime, since $\rho = \alpha^2 T^{00}$ is defined to be the value measured by the normal observers to the spatial slice (not in any rest frame of the fluid), and the averaging of $\bar{\rho}$ across the slice also depends on the gauge. Whilst one might worry that local measures are more sensitive to gauge dependence, particularly in highly dynamical regions, we have nevertheless found these useful in understanding the behaviour of the locally collapsing regions, especially once they decouple from the Hubble flow. Globally averaged measures do not distinguish between regions that are expanding with the Hubble flow and regions that are decoupled and form stable gravitationally bound objects, where the behaviour differs significantly.

Another local quantity that we will track is the compactness of the formed gravitating objects
\begin{equation}
    \mathcal{C}\equiv \frac{G M}{R}~,
\end{equation}
where the limit $\mathcal{C}=1/2$ corresponds to black holes. Directly measuring the mass and radius of the compact objects is non trivial in our inhomogeneous simulations. We search for maximum densities $\rhocosc$ to identify the oscillons and define their surface as the region for which the energy density is $5\%$ of the central value $\rhocosc$. We then calculate their mass and proper volume as the following integrals over regions $\Omega = \{x^i: \rho/\rhocosc>5\% \}$
\begin{align}
    M &=\int_\Omega \dd^3x\sqrt{\gamma} \,\rho~, \label{eq:M}\\
    V &=\int_\Omega \dd^3x\sqrt{\gamma} ~, \label{eq:V}
\end{align}
where $\sqrt{\gamma}$ corresponds to the volume factor of the spatial metric which can be written solely in terms of the conformal factor $\sqrt{\gamma}=\chi^{-3/2}$. We then approximate the radius of the oscillon in terms of its proper volume (in the $t= \mathrm{constant}$ spatial hypersurface of the simulation) as $R=3V^{1/3}/4\pi $.

\section{Growth of overdensities}
\label{sec:results-growth}

\begin{figure}[t!]
    \centering
    \includegraphics[width=\linewidth]{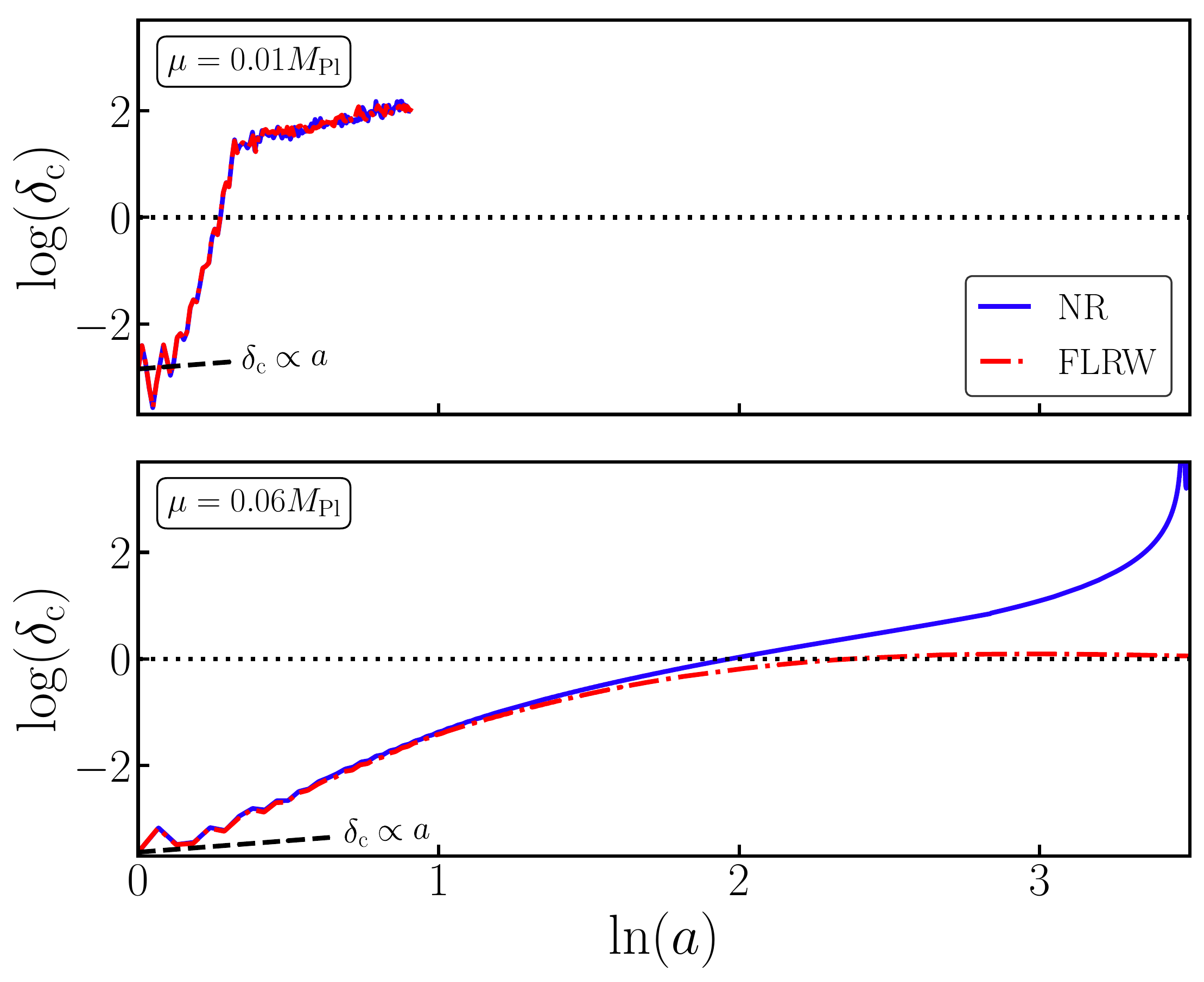}
\caption{Comparison of the growth of perturbations evolved using NR and a spatially-averaged FLRW evolution code for $\mu=0.01\mpl$ (top) and $\mu=0.06\mpl$ (bottom), with initial $\langle\delta\phi^2\rangle\approx 10^{-12}$. The dynamics of overdensities for smaller field models is well-captured with FLRW, whereas for $\mu\gtrapprox 0.06\mpl$ overdensities do not collapse and form oscillons.}
\label{fig:FRW}
\end{figure}

\begin{figure*}[t!]
    \centering
    \includegraphics[width=\linewidth]{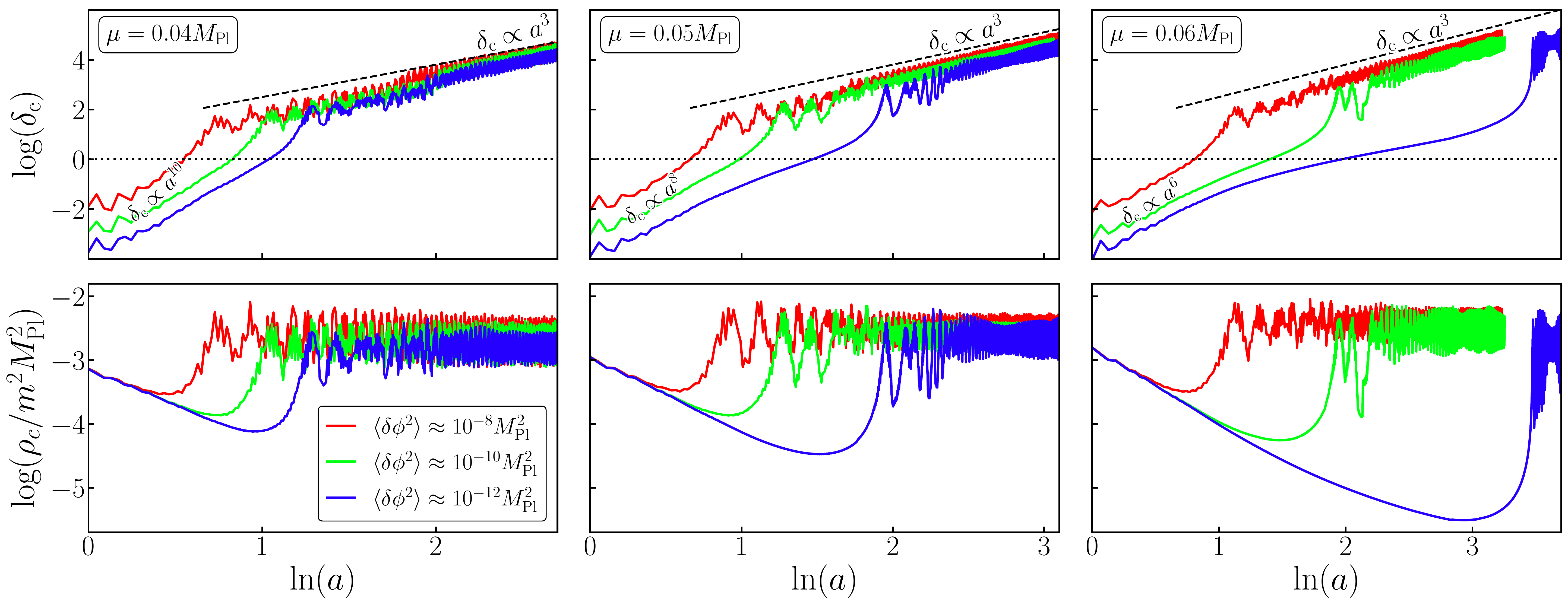}
\caption{Growth of the maximum density contrast $\deltac\equiv \rhoc/\bar{\rho} - 1$ (top panel) and evolution of central densities $\rhoc$ (bottom panel) for a range of potential scales $\mu$ and initial amplitude of fluctuations $\langle \delta\phi^2 \rangle$. We see that the initial rate of growth of perturbations $\deltac$ is largely insensitive to the initial amplitude of the fluctuations $\langle\delta\phi^2\rangle$, with a stronger dependence on the scale of the potential $\mu$. The central value of the overdensities decays until they decouple and turn-around. When they collapse and virialize, oscillons form and the central densities stay at roughly constant values $\rhocosc$. These final central densities are mainly determined by the initial average densities $\rhorehav$ and the growth rate of perturbations during parameteric resonance.}
\label{fig:murho_comparison}
\end{figure*}

In this section we describe the evolution of the density contrast over time, where we plot $\deltac\left(a(t)\right)$ versus time (as measured by the scale factor $a$, which is obtained from the proper volume averaged conformal factor). In Fig.~\ref{fig:FRW}, we compare the results  -- using the same set of parameters -- from the numerical relativity simulations to those that evolve the initial perturbations on a simple spatially-averaged FLRW background. 
For $\langle\delta\phi^2\rangle\approx 10^{-12}\mpl^2$, the disagreement beyond $\mu\gtrapprox 0.06\mpl$ is large enough for FLRW to not capture the collapse of overdensities. 

In the top panel of Fig. \ref{fig:murho_comparison} we study the evolution of the density contrast for different scales of the potential $\mu$ using numerical relativity. We parameterise their initial growth as $\deltac\propto a(t)^\sigma$, where $\sigma$ is a numerical factor that we fit from our simulations. In the cases studied, the fast growth of perturbations with $\sigma\gg 1$ is due to a combination of the parametric and tachyonic resonances together with the gravitational interaction. We find that the rate of growth of perturbations  $\sigma$ during the resonant phase most strongly depends on the scale of the potential $\mu$, whilst it seems largely insensitive to the initial amplitude of the perturbations $\delta_\mathrm{reh}=\deltac(t\reh)$. The values of $\sigma$ are larger for smaller field models (smaller $\mu$), meaning the resonance is stronger and overdensities grow faster. We can relate this to smaller field models exploring more of the tachyonic (concave) part of the potential after inflation. 

We parameterise the evolution of the central values of the overdensity $\rhoc$ as
\begin{equation}\label{eq:rhoc}
    \rhoc\left(a(t)\right) = \frac{\bar{\rho}_\mathrm{reh}}{a^3}\left(\deltaini a^\sigma + 1\right).
\end{equation}
The central density decays as it expands with the background until it decouples from the Hubble flow at the turn-around time $\adec=a(t_\mathrm{dec})$ and starts to collapse. We can estimate $\adec$ as when  $d\rhoc/da=0$, which yields
\begin{align}\label{eq:adec}
    \ln \adec &=\frac{1}{\sigma}\left[\ln\left(\frac{3}{\sigma-3}\right) -\ln\deltaini \right].
\end{align}
These two terms show that $\adec$ decreases for larger initial perturbations $\deltaini$ and faster growth rates (larger $\sigma$), confirming the results in the bottom panel of Fig. \ref{fig:murho_comparison}. Note that we only expect this estimate to be accurate in the case in which the overdensities decouple from the Hubble flow before the growth rate $\sigma$ changes. As the universe expands, the Hubble friction damps the amplitude of the homogeneous oscillations, reducing the resonance. This drives $\sigma$ to smaller values, and thus Eqn. \eqref{eq:adec} only provides a lower bound for $\adec$. 

After decoupling, overdensities follow their own field and gravitational dynamics, with a characteristic timescale for the collapse that seems to be mostly dependent on $\rhocdec = \rhoc(\adec)$ via the free-fall timescale. During the collapse, the central densities $\rhoc$ bounce back to larger values until the overdensities virialize and saturate at a roughly constant density $\rhocosc=\rhoc(a_\mathrm{osc})$, forming stable scalar field configurations -- oscillons. This can be seen in Fig. \ref{fig:murho_comparison}, with the central density oscillating about some roughly constant value and giving rise to a density contrast scaling as $\deltac\propto a^3$ due to the matter dominated average density decay $\bar{\rho}\propto a^{-3}$. It should be noted that, while in this paper we have collectively denoted pseudo-stable real scalar field configuration with the term \textit{oscillon}, when the pseudo-stability is due to gravitational effects, these configurations are more properly known as \textit{oscillatons} \cite{Kaup:1968zz,Ruffini:1969qy,Hogan:1988mp,Seidel:1991zh,Liddle:1992fmk,Seidel:1993zk,Kolb:1993zz,Kolb:1993hw,Urena-Lopez:2002ptf,Urena-Lopez:2001zjo,Alcubierre:2003sx,Valdez-Alvarado:2011xnr,Urena-Lopez:2012udq,Krippendorf:2018tei,Ikeda:2017qev}.

\section{Oscillon formation}
\label{sec:results-endpoint}

We expect that the interplay between the overdensity growth rate and the Hubble expansion rate dictates the final values of the central densities $\rhocosc$, which in turn determine the properties of the oscillons that are formed. These depend on two factors:
\begin{itemize}
    \item[(i)] The central density decays until decoupling at $\adec$: The smaller $\adec$, the larger $\rhocosc$.
    \item[(ii)] Initially, the central density is approximately $\bar{\rho}\reh$: The larger $\bar{\rho}\reh$, the larger $\rhocosc$.
\end{itemize}

We can check (i) by looking at the results for a given model $\mu$, as those share the same initial density $\bar{\rho}\reh$. The bottom panels of Fig. \ref{fig:murho_comparison} confirm that overdensities that decouple the earlier (smaller $\adec$) have larger final densities $\rhocosc$.

Checking (ii) is more difficult as both $\bar{\rho}\reh$ and $\adec$ play a role in a non-trivial manner.  Larger field models start from larger initial densities $\rhorehav$, but experience weaker resonances with slower growth rates $\sigma$, resulting in longer decoupling times $\adec$. There is therefore competition between these effects, such that larger initial $\rhorehav$ can balance the extra decay from larger $\adec$, resulting in $\rhocosc$ growing with  $\mu$.\\

\begin{figure}[t!]
    \centering
    \includegraphics[width=\linewidth]{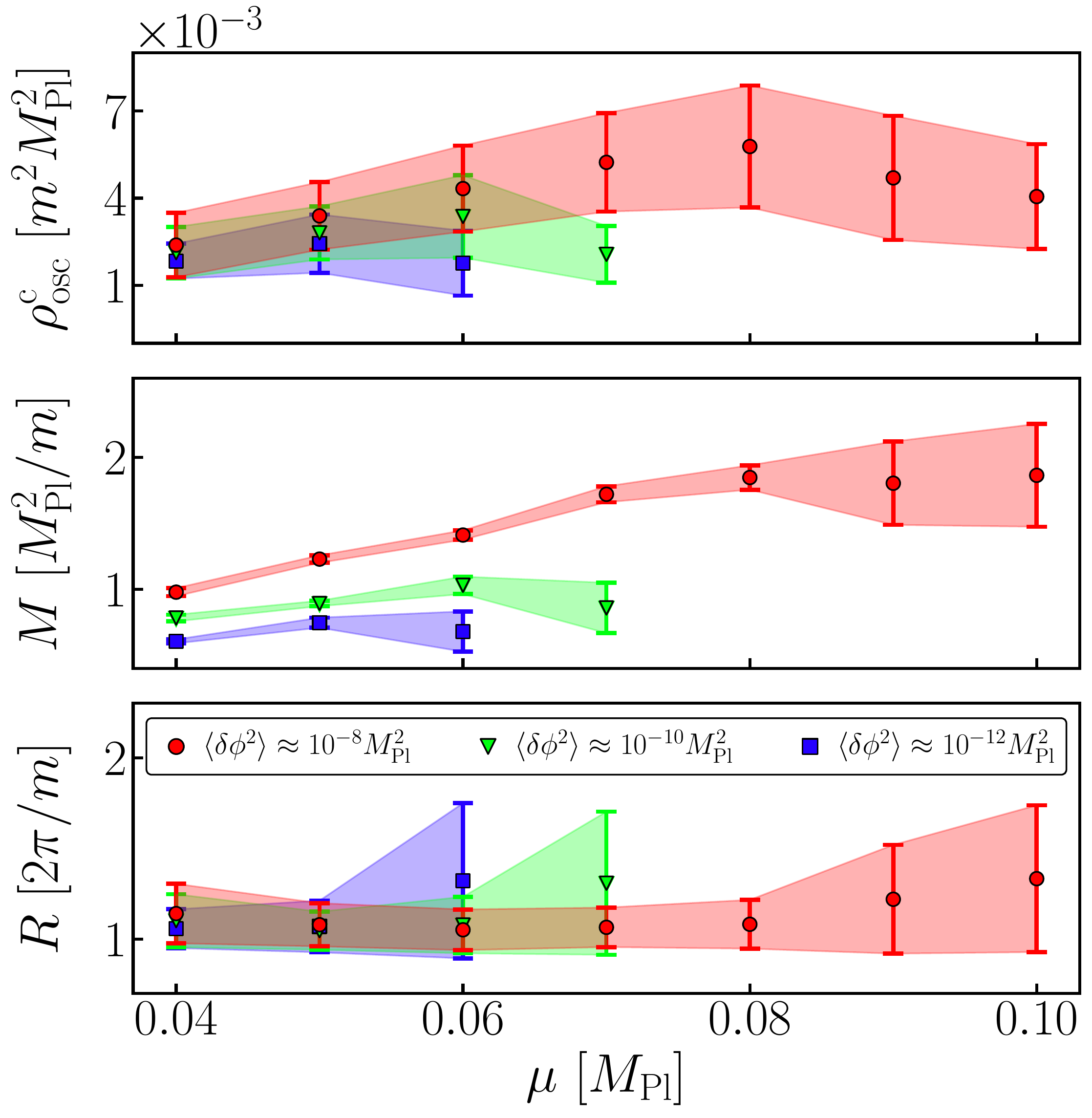}
\caption{Central energy densities, masses and radii of oscillons versus scale of the potential $\mu$, for different initial fluctuations $\langle \delta\phi^2 \rangle$. The central densities and masses increase until they peak at a critical $\mu$, whilst the radii remain at a roughly constant value $R\approx 2\pi/m$. The values and error bars are estimated by computing the average and standard deviation of the evolution of $\rhocosc$, $M$ and $R$ over the last $0.2$ e-foldings of the simulation.}
\label{fig:MR_comparison}
\end{figure}

Let us assume that the oscillon central density $\rhocosc$ is approximately determined by the densities at decoupling $\rhocdec$, which scale via Eqns. (\ref{eq:rhoc}-\ref{eq:adec}) as
\begin{equation}
    \rhocosc\propto \rhocdec=\rhorehav \deltaini^{1-x} x^{-x} (1-x)^{-(1-x)} ~,
\end{equation}
where we have defined $x\equiv (\sigma-3)/\sigma$ for convenience. Since $\rhocosc\propto \rhorehav$, then $\rhocosc\propto m^2\mu^2$. The function $\deltaini^{1-x}x^{-x}(1-x)^{-(1-x)}$ has a strong dependence on $\sigma$, which itself is very sensitive to the scale of the inflationary model $\mu$. We therefore expect that $\rhocosc$ will grow with $\mu$ if $\sigma \gg 3$, whereas the expansion will overtake (and $\rhocosc$ decay with $\mu$) as $\sigma\rightarrow 3$. Based on these arguments, the oscillon density $\rhocosc$ should peak at some value $(\mu,\langle \delta\phi^2 \rangle)$. We confirm this in the top panel of  Fig. \ref{fig:MR_comparison}, where an initially increasing trend in the density can be observed up to  a critical $(\mu,\langle \delta\phi^2 \rangle)$, after which $\rhocosc$ decays.

Once the location of final central densities $\rhocosc$ is identified, we can compute other oscillon properties such as their mass and volume using Eqns. (\ref{eq:M}-\ref{eq:V}). The mass is related to the final central density of the oscillon, which depends on the details of the formation process as discussed above. It therefore also shows a peak for each model value $\mu$, as shown in the middle panel of Fig. \ref{fig:MR_comparison}. The radius $R$, on the other hand, is mainly determined by the mass $m$ around the reheating minimum 
%which is related to the period of the oscillation 
as $R\approx 2\pi /m$, bottom panel of Fig. \ref{fig:MR_comparison}. Hence the peak in the resulting compactnesses shown in Fig. \ref{fig:oscillon}. The values and error bars in Figs. \ref{fig:oscillon} and \ref{fig:MR_comparison} are estimated by computing the average and standard deviation of these properties over the last $0.2$ e-folds of the simulation.

\section{Conclusions}
\label{sec:conclusion}

In this work we have shown that perturbations from the most minimal preheating scenario -- that of resonance in a single field inflationary model that is consistent with current CMB observations -- can grow and collapse to form oscillons with compactnesses $\mathcal{C}\sim \mathcal{O}(10^{-3})$. In particular, we have observed that the interplay between the overdensity growth and Hubble expansion rates results in a maximum compactness that depends on $\langle\delta\phi^2\rangle$ and can reach $\mathcal{C}\approx 10^{-2}$ in the regime we have studied.

This is not sufficient to drive the oscillons to collapse to black holes, and it seems likely that increasing the initial amplitudes $\langle\delta\phi^2\rangle$ is the only way one could reach black hole formation\footnote{Previous works finding black hole formation \cite{Helfer:2016ljl, Michel:2018nzt, Widdicombe:2018oeo, Muia:2019coe, Nazari:2020fmk} have often started with highly compact initial perturbations, considered a higher degree of symmetry, or used smaller volumes comparable to the size of the overdensity itself, which may have prevented radiation from being efficient during the collapse (due to the use of periodic boundary conditions for example).}. However, that would take us outside of the values expected from our minimal inflationary preheating scenario as $\deltaini\rightarrow 1$. Hence, we conclude that inflationary preheating is not enough to form primordial black holes without an additional enhancement mechanism of the perturbations.

Constraints on the scalar power spectrum fix the mass of the field $m\approx 10^{-5}\mpl$ and set the size and mass of these relics to be $R\approx 10^{-33} \,\mathrm{km}$ and $M\approx 10^{-3} \,\mathrm{g}$. They are not suitable dark matter candidates as they will decay into standard model particles if they have additional couplings\footnote{In the case of coupling to photons, oscillons could decay in electromagnetic radiation \cite{Levkov:2020txo,Chung-Jukko:2023cow}.}. However, they can leave an observational imprint on the stochastic background of gravitational waves during their formation process and subsequent evolution. We have observed that this can appear rather chaotic, with some oscillons fragmenting into multiple objects. Naively, we expect the radiation from such processes to increase with the compactness of oscillons, as was observed in the context of oscillaton collisions \cite{Helfer:2018vtq}, but we leave a systematic study of the precise gravitational-wave spectrum for future work.

\section*{Acknowledgements}

We would like to thank David Alonso, Pedro Ferreira, Cristian Joana, Robyn Munoz, Martin Rey, Evangelos Sfakianakis and Francisco Torrenti for helpful conversations. We thank the GRChombo collaboration (\href{www.grchombo.org}{www.grchombo.org}) for their support and code development work. JCA acknowledges funding from the Beecroft Trust and The Queen’s College via an extraordinary Junior Research Fellowship (eJRF). KC acknowledges funding from the UKRI Ernest Rutherford Fellowship (grant number ST/V003240/1). FM is funded by a UKRI/EPSRC Stephen Hawking fellowship, grant reference EP/T017279/1, partially supported by the STFC consolidated grant ST/P000681/1 and funded by a G-Research grant for postdocs in quantitative fields. 
% For the purpose of Open Access, the author has applied a CC BY public copyright licence to any Author Accepted Manuscript version arising from this submission.

This work was performed using the DiRAC@Durham facility managed by the Institute for Computational Cosmology on behalf of the STFC DiRAC HPC Facility (www.dirac.ac.uk) under DiRAC RAC13 Grant ACTP238 and DiRAC RAC15 Grant ACTP316. The equipment was funded by BEIS capital funding via STFC capital grants ST/P002293/1, ST/R002371/1 and ST/S002502/1, Durham University and STFC operations grant ST/R000832/1. DiRAC is part of the National e-Infrastructure.

\bibliography{mybib}

\newpage
\clearpage

\appendix

\counterwithin{figure}{section}

\section{Numerical methodology}
\label{app:numerics}

\begin{figure}[t!]
    \centering
\includegraphics[width=\linewidth]{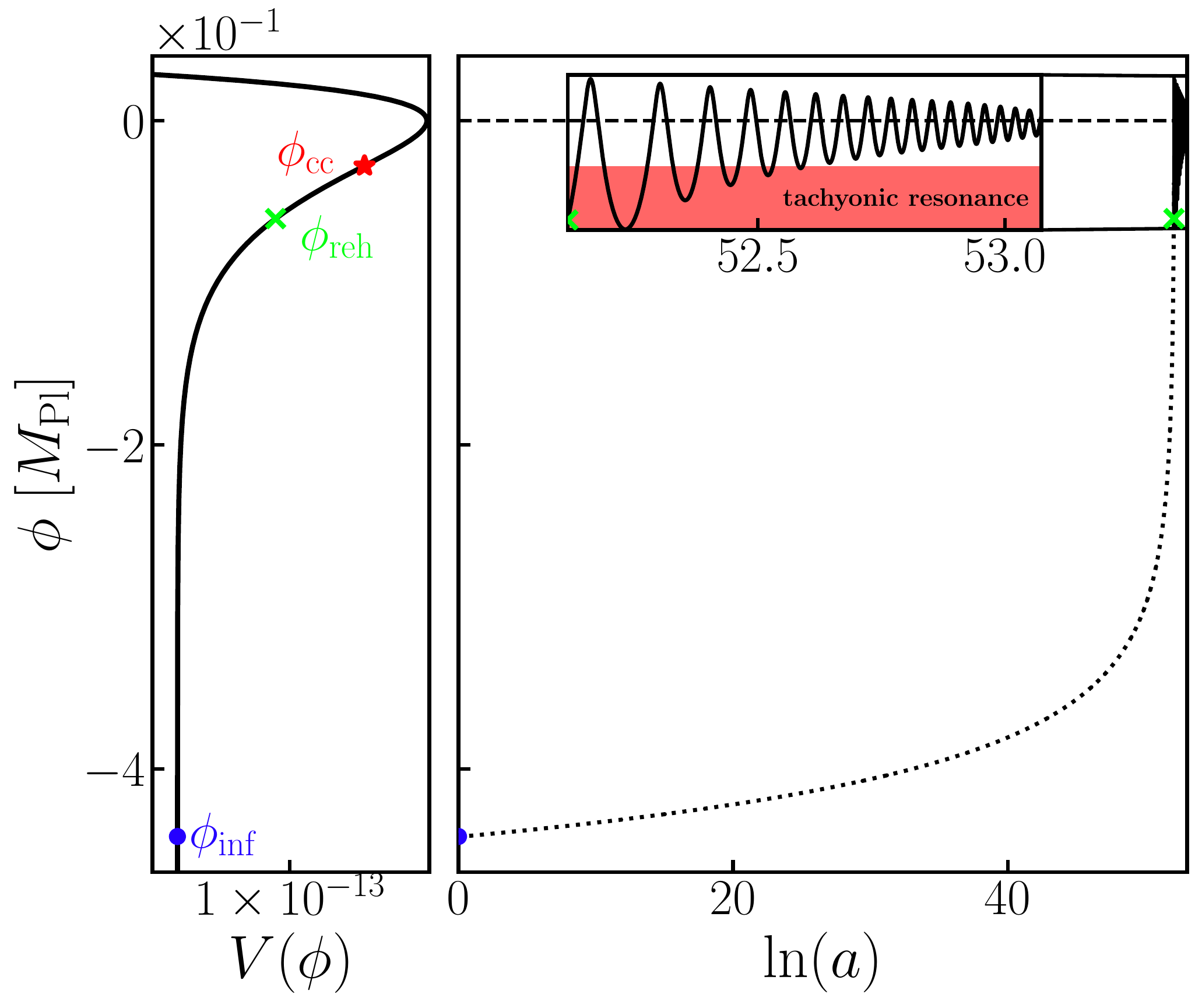}
\caption{Solution of the homogeneous case that is used to set the initial values of the field in our simulations. The left panel shows the potential (seen `on its side') and the right panel shows the evolution of the field over time within it. Our simulations start at $\phi_{\rm reh}$, where inflation ends and the reheating phase begins. The shape of the potential changes from concave to convex at $\phi_\mathrm{cc}=-\mu\ln 2$. In the inset panel we see that the initial oscillations in the field probe the tachyonic part of the potential (shaded in red). At later times the resonance is purely parametric.}
\label{fig:homogeneous}
\end{figure}

Here we describe the methodology used to obtain initial conditions that are consistent with constraints on inflation from the CMB. In the absence of inhomogeneities, we take the spacetime to be well described by the (flat) FLRW metric
\begin{equation}
    ds^2 = -dt^2 + a(t)^2(dx^2 + dy^2 + dz^2)~,
\end{equation}
where $a(t)$ is the scale factor that evolves as
\begin{equation}\label{eq:flrw}
    \frac{\ddot{a}}{a} = -\frac{1}{3\mpl^2}\left(\frac{1}{2}\dot{\phi}^2 -V(\phi)\right)~.
\end{equation}
The inflaton is driven by the Klein-Gordon equation
\begin{equation}\label{eq:kg}
    \ddot{\phi} + 3H\dot{\phi} + V'(\phi) = 0~,
\end{equation}
where $H\equiv \dot{a}/a$ is the Hubble parameter. We choose the initial value of the scalar field $\phi_\mathrm{inf}=\phi(t_\mathrm{inf})$ by requiring inflation to last $\mathcal{N}\approx 50$ e-folds, see Fig. \ref{fig:homogeneous}. We solve the FLRW evolution equations and define the end of inflation (and beginning of reheating) when $\ddot{a}(t\reh)=0$. At this point, we extract the values of the scalar field and momentum $\phi\reh = \phi(t\reh)$ and $\dot{\phi}\reh= \dot{\phi}(t\reh)$, which will be used as the homogeneous components of the initial conditions in our simulations. We set the size of our numerical domain to be approximately the Hubble length at the end of inflation, $L\approx H(t\reh)^{-1}$. We factor out the mass $m$ in Eqn. \eqref{eq:potential} and measure everything in units of $m$, which is fixed by constraints on the scalar power spectrum.

To obtain the perturbations we follow closely the methods of ~\cite{Figueroa:2020rrl}. Assuming a random Gaussian field, the spectrum of sub-horizon scalar perturbations at the end of inflation is determined by quantum vacuum fluctuations
\begin{align}
\label{eq:PertSpectrum}
\mathcal{P}(k) &= \frac{\lambda}{2 a\reh^2 \omega_k^2} \,, \\
\langle \delta\phi_\mathbf{k} \delta\phi_{\mathbf{k}'} \rangle &= (2\pi)^3 \mathcal{P}(k) \delta(\mathbf{k}-\mathbf{k}') \,,
\end{align}
where $\omega_k = \sqrt{k^2 + a\reh^2 V''(\phi\reh)}$ and we will take the initial scale factor to be $a\reh=1$. The spectrum in Eq.~\eqref{eq:PertSpectrum} determines the variance of the field perturbations through the relation
\begin{equation}
\langle \delta\phi^2 \rangle = \int d\log k \, \frac{k^3}{2 \pi^2} \mathcal{P}(k) \,,
\end{equation}
where we study the impact of varying $\lambda$ by a few orders of magnitude.

In order to mimic such a spectrum in the simulations, we consider a box of physical size $L$ and $N$ gridpoints in each of the three spatial directions, without adaptive mesh refinement initially. Then, the grid contains $N^3$ points labeled as $\mathbf{n} = (n_1, n_2, n_3)$, with $n_i = \lbrace 0, \dots, N-1\rbrace$ ($i = 1, 2, 3$). The physical distance between adjacent gridpoints is $\Delta x = L/N$. The reciprocal lattice is then given by $\tilde{\mathbf{n}} = (\tilde{n}_1, \tilde{n}_2, \tilde{n}_3)$, where $\tilde{n}_i = \lbrace -N/2+1, -N/2, \dots, -1, 0, 1, \dots, N/2-1, N/2 \rbrace$ ($i = 1,2,3$), so that the wavenumber corresponding to each point on the reciprocal lattice is $\mathbf{k}(\mathbf{\tilde{n}}) = 2 \pi \mathbf{\tilde{n}}/(N \Delta x)$. Note that every continuous function $f(x)$ becomes a function of $\mathbf{n}$ on the grid, and can be Fourier transformed as $f(\mathbf{n}) = \frac{1}{N^3} \sum_{\mathbf{\tilde{n}}} e^{-i \frac{2 \pi}{N} \mathbf{n} \cdot \mathbf{\tilde{n}}} f(\mathbf{\tilde{n}})$. We mimic the spectrum in Eq.~\eqref{eq:PertSpectrum}, building the pointwise perturbations at each point in the reciprocal lattice as in~\cite{Figueroa:2020rrl}
\begin{align}
\delta \phi(\tilde{\mathbf{n}}) =& \frac{1}{\sqrt{2}} \left(|\delta\phi^{\rm(l)} (\tilde{\mathbf{n}})| e^{i \theta^{\rm(l)} (\tilde{\mathbf{n}})} + |\delta\phi^{\rm(r)} (\tilde{\mathbf{n}})| e^{i \theta^{\rm(r)} (\tilde{\mathbf{n}})}\right) \nonumber \\
m \,\delta \dot\phi(\tilde{\mathbf{n}}) =& \frac{1}{a\reh} \left[\frac{i \omega_k}{\sqrt{2}} \left(|\delta\phi^{\rm(l)} (\tilde{\mathbf{n}})| e^{i \theta^{\rm(l)} (\tilde{\mathbf{n}})} - |\delta\phi^{\rm(r)} (\tilde{\mathbf{n}})| e^{i \theta^{\rm(r)} (\tilde{\mathbf{n}}) }\right)\right]  \nonumber \\
&- H \, \delta\phi(\tilde{\mathbf{n}}) \,. \nonumber
\end{align}
We draw $\theta^{(l)}(\tilde{\mathbf{n}})$ and $\theta^{(r)}(\tilde{\mathbf{n}})$ from a uniform distribution in $[0, 2\pi)$, and we generate the amplitudes $\delta\phi^{(l)}(\tilde{\mathbf{n}})$, and $\delta\phi^{(r)}(\tilde{\mathbf{n}})$ from a Rayleigh distribution with expected square amplitude given by
\begin{equation}
|\delta\phi(\mathbf{\tilde{n})}|^2 = \left(\frac{N}{\Delta x}\right)^3 \frac{\lambda}{2 a\reh^2 \sqrt{(k(\mathbf{\tilde{n}}))^2 + a\reh^2 V''(\phi\reh)}} \,.
\end{equation}

Note that the maximum wavenumber that we are able to capture is $\pi/\Delta x$, corresponding to the minimum wavelength equal to the distance between two adjacent gridpoints in the box. However, in order to avoid including these underresolved modes in the initial conditions, we mask the spectrum with an ultraviolet cutoff $k_\mathrm{UV}=4 k_\mathrm{IR}=4(2\pi/L)$. Including frequencies that experience parametric resonance is clearly important, but provided such modes are included, the precise value of the cut off does not lead to significantly different behaviour.

We then evolve the BSSN system of equations \cite{Nakamura:1987zz,Shibata:1995we,Baumgarte:1998te} together with the Einstein-Klein-Gordon equation \eqref{eq:EKG_cov} decomposed into two first order equations, as
\begin{align}
    \label{eq:EKG1}\partial_t \phi  =& \alpha\Pi +\beta^i\partial_i\phi~, \\
    \partial_t \Pi =& \beta^i\partial_i\Pi + \alpha\gamma^{ij}(\partial_i \partial_j \phi +\partial_i\phi \partial_j \alpha) \nonumber\\
    &+ \alpha\left(K\Pi - \gamma^{ij}\Gamma^k_{ij}\partial_k\phi - \frac{dV}{d\phi}\right) . \label{eq:EKG2}
\end{align}

\section{Impact of backreaction:\\ Numerical relativity versus FLRW}
\label{app:FLRW}

Comparing the evolution in FLRW simulations and our fully non linear simulations is inherently difficult due to the potential for gauge ambiguities (in particular, the absence of a well defined background evolution for the inhomogeneous case). However, to get an idea of the potential differences, we adapt \textsc{grchombo} in a simple manner to evolve the scalar field equations of motion (\ref{eq:EKG1}-\ref{eq:EKG2}) in a spatially-constant FLRW background. We evolve cosmic time slices, enforcing isotropy and homogeneity in the metric sector and fixing the gauge variables $\alpha=1$ and $\beta^i=0$. The BSSN equations of motion are then simplified to
\begin{align}
    \partial_t\chi &= \frac{2}{3}\chi K \\
    \partial_t K &= \frac{1}{3} K^2 + \frac{1}{2\mpl^2} (\bar{\rho} + \bar{S})~,
\end{align}
where $\bar{\rho}$ and $\bar{S}$ correspond to the volume-averaged energy density and trace of the spatial stress components of the stress-energy tensor.  We initially set $\chi = 1$ and enforce that the trace of the extrinsic curvature tensor satisfies the Hamiltonian (Friedmann) constraint $K^2=3\bar{\rho}/\mpl^2$. Noting that $\chi=a^{-2}$, $K=-3H$ and $\bar{S}=3\mathrm{p}$, these two equations are equivalent to solving $H=\dot{a}/a$ and $\dot{H}=-H^2-(\bar{\rho} + \bar{S})/6\mpl^2$, respectively, which in turn is the usual Friedmann equation $\ddot{a}/a=-(\bar{\rho} + 3\mathrm{p})/6\mpl^2$.\\

Results for two scales $\mu$ of the potential are shown in Fig. \ref{fig:FRW}. In agreement with previous studies \cite{Kou:2019bbc}, we find that for smaller field models the dynamics can be well captured by a spatially-averaged FLRW code. For larger field models, on the other hand, taking into account gravitational effects can be crucial for the collapse of overdensities and formation of oscillons.

\vspace{20pt}
\section{Summary of simulations and convergence tests}
\label{app:testing}

\begin{figure}[b]
    \centering
    % \vspace{-5pt}
\includegraphics[width=\linewidth]{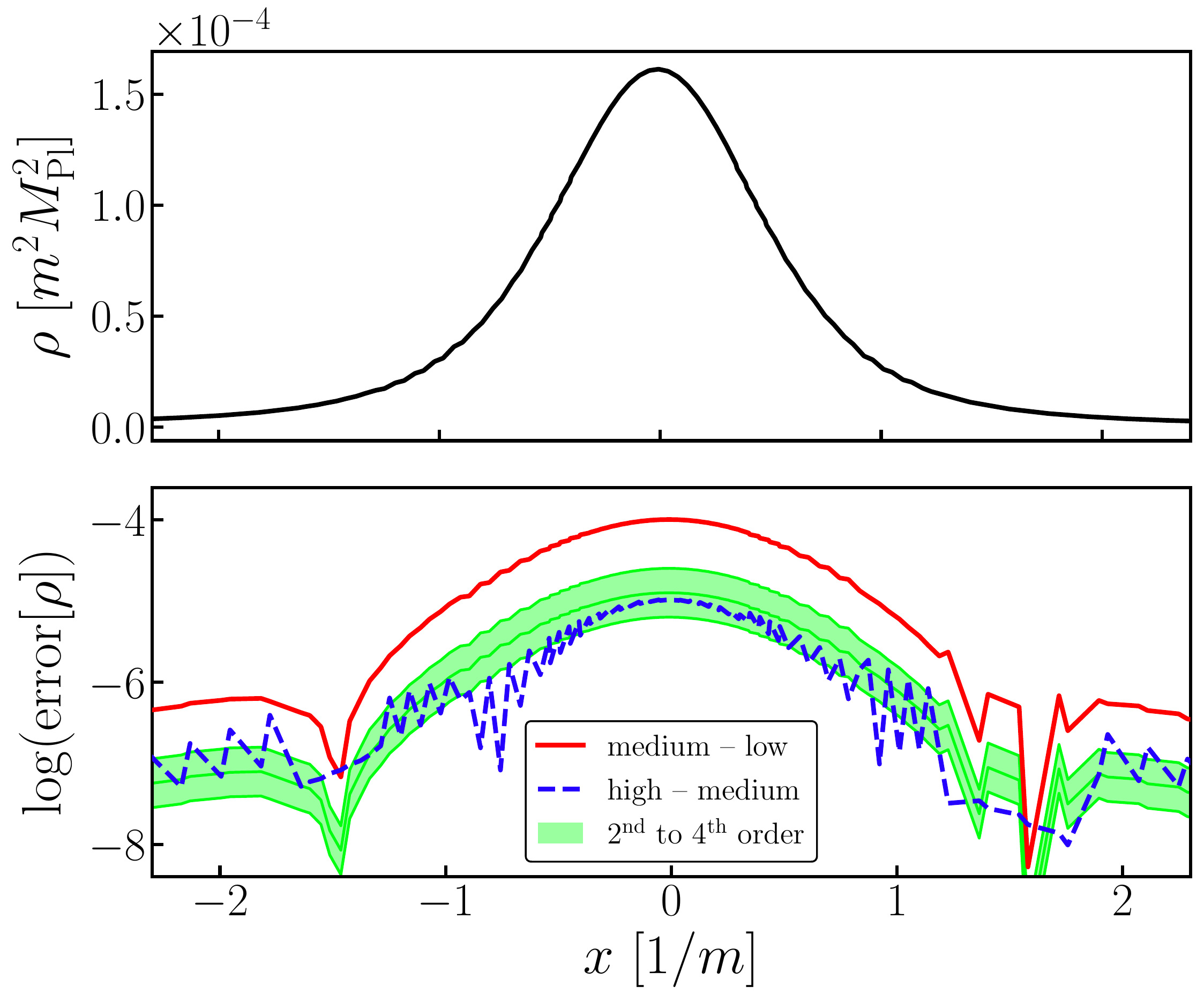}
\caption{Convergence test of an oscillon density profile  (top panel) formed for $\mu=0.1\mpl$ and $\langle\delta\phi^2\rangle\approx 10^{-8}\mpl^2$, consistent with $2^\mathrm{nd}$ to $4^\mathrm{th}$ order convergence, as expected. The error in the bottom panel is calculated by subtracting the density profiles $\rho$ for low ($N=64$), medium ($N=128$) and high ($N=256$) resolutions, with $5$ levels of adaptive mesh refinement.}
\label{fig:convergence_rho}
\end{figure}

\begin{table*}[t!]
\centering
\begin{tabular}{|c|c|c|c|c|c|c|c|c|}
\hline
$\quad\mu~[\mpl]\quad$ & $\quad m~[\mpl]\quad$   & $\phi_\mathrm{inf}~[\mpl]$             &  $\quad H_\mathrm{inf}~[m]\quad$ & $\phi\reh~[\mpl]$ &  $\dot{\phi}\reh~[m \mpl]$ & $\quad H\reh~[m]\quad$ & $\quad H^{-1}\reh~[m^{-1}]\quad$  \\ \hline

$0.1$ & $1.50501\ee{-5}$ & $-9.2125\ee{-1}$ & $4.08208\ee{-2}$ & $-1.32906\ee{-1}$ & $5.19917\ee{-2}$ & $3.67637\ee{-2}$ & $27.2008$ \\ \hline

$0.09$ & $1.50536\ee{-5}$ & $-8.4807\ee{-1}$ & $3.67394\ee{-2}$ & $-1.21893\ee{-1}$ & $4.72136\ee{-2}$ & $3.33851\ee{-2}$ & $29.9535$ \\ \hline

$0.08$ & $1.50571\ee{-5}$ & $-7.7266\ee{-1}$ & $3.26578\ee{-2}$ & $-1.10495\ee{-1}$ & $4.2354\ee{-2}$ & $2.99488\ee{-2}$ & $33.3903$ \\ \hline

$0.07$ & $1.50604\ee{-5}$ & $-6.94755\ee{-1}$ & $2.8576\ee{-2}$ & $-9.86792\ee{-2}$ & $3.74094\ee{-2}$ & $2.64525\ee{-2}$ & $37.8037$ \\ \hline

$0.06$ & $1.50636\ee{-5}$ & $-6.13988\ee{-1}$ & $2.4494\ee{-2}$ & $-8.64102\ee{-2}$ & $3.23761\ee{-2}$ & $2.28934\ee{-2}$ & $43.6807$\\ \hline

$0.05$ & $1.50666\ee{-5}$ & $-5.29877\ee{-1}$ & $2.04119\ee{-2}$ & $-7.36454\ee{-2}$ & $2.72499\ee{-2}$ & $1.92686\ee{-2}$ & $51.898$ \\ \hline

$0.04$ & $1.50696\ee{-5}$ & $-4.41745\ee{-1}$ & $1.63297\ee{-2}$ & $-6.03334\ee{-2}$ & $2.20256\ee{-2}$ & $1.55744\ee{-2}$ & $64.2078$ \\ \hline
\end{tabular}
 \caption{\textbf{Summary of simulations:} We freely choose the scale of the potential $\mu$, whilst the mass of the field $m$ is fixed by the scalar index measurements from the Planck Collaboration \cite{Planck:2018jri}. The values of the scalar field $\phi_\mathrm{inf}$ and Hubble parameter $H_\mathrm{inf}$ at the beginning of inflation would result in $\ln(a)\approx 50$ e-folds until the start of reheating with $\phi_\mathrm{reh}$ and $H_\mathrm{reh}$. We choose the size of the simulation box to be of order the Hubble length at the end of inflation $L=\mathcal{O}(H\reh^{-1})=64m^{-1}$.}
\label{fig:table}
\end{table*}

A summary table of the simulation parameters is given in table \ref{fig:table}. We have chosen a box size of length $L=64m^{-1}$ for all simulations, which would correspond to the largest Hubble length studied, corresponding to $\mu=0.04\mpl$. For convergence testing, our high, medium and low resolution runs have $N^3=\lbrace 256^3,128^3,64^3\rbrace$ number of coarse grid points respectively, in addition to $5$ level of refinements. A convergence test of one of the oscillon density profiles (for the largest $\mu=0.1\mpl$ case) is shown in Fig. \ref{fig:convergence_rho}. This shows clear $2^\mathrm{nd}$ to $4^\mathrm{th}$ order convergence, as expected from the finite difference schemes in the initial condition and evolution codes.

\end{document}